\pdfoutput=1

\documentclass[12pt,a4paper]{article}

\usepackage{ifthen} 
\newboolean{pdflatex}
\setboolean{pdflatex}{true} 

\newboolean{articletitles}
\setboolean{articletitles}{true} 

\newboolean{uprightparticles}
\setboolean{uprightparticles}{false} 

\usepackage{longtable} 

\usepackage{microtype}
\usepackage{lineno}  
\usepackage{xspace} 

\usepackage{graphicx}  
\usepackage{color}
\usepackage{colortbl}
\graphicspath{{./figs/}} 

\usepackage{amsmath} 
\usepackage{amssymb}
\usepackage{amsfonts}
\usepackage{upgreek} 

\newcommand*\patchAmsMathEnvironmentForLineno[1]{%
\expandafter\let\csname old#1\expandafter\endcsname\csname #1\endcsname
\expandafter\let\csname oldend#1\expandafter\endcsname\csname
end#1\endcsname
 \renewenvironment{#1}%
   {\linenomath\csname old#1\endcsname}%
   {\csname oldend#1\endcsname\endlinenomath}%
}
\newcommand*\patchBothAmsMathEnvironmentsForLineno[1]{%
  \patchAmsMathEnvironmentForLineno{#1}%
  \patchAmsMathEnvironmentForLineno{#1*}%
}
\AtBeginDocument{%
\patchBothAmsMathEnvironmentsForLineno{equation}%
\patchBothAmsMathEnvironmentsForLineno{align}%
\patchBothAmsMathEnvironmentsForLineno{flalign}%
\patchBothAmsMathEnvironmentsForLineno{alignat}%
\patchBothAmsMathEnvironmentsForLineno{gather}%
\patchBothAmsMathEnvironmentsForLineno{multline}%
}

\usepackage{hyperref}    
\usepackage[all]{hypcap} 

\def\lhcb {\mbox{LHCb}\xspace}
\def\ux85 {\mbox{UX85}\xspace}

\ifthenelse{\boolean{uprightparticles}}%
{

 \def\Ppsi        {\ensuremath{\uppsi}\xspace}

 \def\PDelta      {\ensuremath{\Delta}\xspace}                 
 \def\PXi      {\ensuremath{\Xi}\xspace}                 
 \def\PLambda      {\ensuremath{\Lambda}\xspace}                 
 \def\PSigma      {\ensuremath{\Sigma}\xspace}                 
 \def\POmega      {\ensuremath{\Omega}\xspace}                 
 \def\PUpsilon      {\ensuremath{\Upsilon}\xspace}                 
 
 \def\PB      {\ensuremath{\mathrm{B}}\xspace}                 
                  
 \def\PD      {\ensuremath{\mathrm{D}}\xspace}

 \def\PJ      {\ensuremath{\mathrm{J}}\xspace}                 
 \def\PK      {\ensuremath{\mathrm{K}}\xspace}

 \def\Pb      {\ensuremath{\mathrm{b}}\xspace}                 
 \def\Pc      {\ensuremath{\mathrm{c}}\xspace}

 \def\Pi      {\ensuremath{\mathrm{i}}\xspace}

}
{

 \def\Ppsi        {\ensuremath{\psi}\xspace}                 
                  
 \mathchardef\PDelta="7101
 \mathchardef\PXi="7104
 \mathchardef\PLambda="7103
 \mathchardef\PSigma="7106
 \mathchardef\POmega="710A
 \mathchardef\PUpsilon="7107
                  
 \def\PB      {\ensuremath{B}\xspace}                 
                  
 \def\PD      {\ensuremath{D}\xspace}

 \def\PJ      {\ensuremath{J}\xspace}                 
 \def\PK      {\ensuremath{K}\xspace}

 \def\Pb      {\ensuremath{b}\xspace}                 
 \def\Pc      {\ensuremath{c}\xspace}

 \def\Pi      {\ensuremath{i}\xspace}

}

\def\cquark    {\ensuremath{\Pc}\xspace}

\def\bquark    {\ensuremath{\Pb}\xspace}

\def\pipi  {\ensuremath{\pion^+\pion^-}\xspace}

\def\kaon  {\ensuremath{\PK}\xspace}
  \def\Kbar  {\kern 0.2em\overline{\kern -0.2em \PK}{}\xspace}

\def\Kz    {\ensuremath{\kaon^0}\xspace}
\def\Kzb   {\ensuremath{\Kbar^0}\xspace}
\def\KzKzb {\ensuremath{\Kz \kern -0.16em \Kzb}\xspace}
\def\Kp    {\ensuremath{\kaon^+}\xspace}
\def\Km    {\ensuremath{\kaon^-}\xspace}

\def\KpKm  {\ensuremath{\Kp \kern -0.16em \Km}\xspace}

\def\Kstar   {\ensuremath{\kaon^*}\xspace}

  \def\Dbar    {\kern 0.2em\overline{\kern -0.2em \PD}{}\xspace}
\def\D       {\ensuremath{\PD}\xspace}

\def\Dz      {\ensuremath{\D^0}\xspace}
\def\Dzb     {\ensuremath{\Dbar^0}\xspace}
\def\DzDzb   {\ensuremath{\Dz {\kern -0.16em \Dzb}}\xspace}
\def\Dp      {\ensuremath{\D^+}\xspace}
\def\Dm      {\ensuremath{\D^-}\xspace}

\def\DpDm    {\ensuremath{\Dp {\kern -0.16em \Dm}}\xspace}

\def\Bbar    {\ensuremath{\kern 0.18em\overline{\kern -0.18em \PB}{}}\xspace}

\def\jpsi     {\ensuremath{{\PJ\mskip -3mu/\mskip -2mu\Ppsi\mskip 2mu}}\xspace}

  \def\Y#1S{\ensuremath{\PUpsilon{(#1S)}}\xspace}

\def\L {\ensuremath{\PLambda}\xspace}
\def\Lbar {\ensuremath{\kern 0.1em\overline{\kern -0.1em\PLambda}}\xspace}

\def\BR         {\BF}

\def\to                 {\ensuremath{\rightarrow}\xspace}

\def\AT#1     {\ensuremath{A_{\mathrm{T}}^{#1}}\xspace}           

\def\C#1      {\ensuremath{\mathcal{C}_{#1}}\xspace}                       
\def\Cp#1     {\ensuremath{\mathcal{C}_{#1}^{'}}\xspace}                    
\def\Ceff#1   {\ensuremath{\mathcal{C}_{#1}^{\mathrm{(eff)}}}\xspace}        
\def\Cpeff#1  {\ensuremath{\mathcal{C}_{#1}^{'\mathrm{(eff)}}}\xspace}       
\def\Ope#1    {\ensuremath{\mathcal{O}_{#1}}\xspace}                       
\def\Opep#1   {\ensuremath{\mathcal{O}_{#1}^{'}}\xspace}                    

\newcommand{\tev}{\ensuremath{\mathrm{\,Te\kern -0.1em V}}\xspace}
\newcommand{\gev}{\ensuremath{\mathrm{\,Ge\kern -0.1em V}}\xspace}
\newcommand{\mev}{\ensuremath{\mathrm{\,Me\kern -0.1em V}}\xspace}
\newcommand{\kev}{\ensuremath{\mathrm{\,ke\kern -0.1em V}}\xspace}
\newcommand{\ev}{\ensuremath{\mathrm{\,e\kern -0.1em V}}\xspace}
\newcommand{\gevc}{\ensuremath{{\mathrm{\,Ge\kern -0.1em V\!/}c}}\xspace}
\newcommand{\mevc}{\ensuremath{{\mathrm{\,Me\kern -0.1em V\!/}c}}\xspace}
\newcommand{\gevcc}{\ensuremath{{\mathrm{\,Ge\kern -0.1em V\!/}c^2}}\xspace}
\newcommand{\gevgevcccc}{\ensuremath{{\mathrm{\,Ge\kern -0.1em V^2\!/}c^4}}\xspace}
\newcommand{\mevcc}{\ensuremath{{\mathrm{\,Me\kern -0.1em V\!/}c^2}}\xspace}

\def\mum  {\ensuremath{\,\upmu\rm m}\xspace}

\newcommand{\chisq}{\ensuremath{\chi^2}\xspace}

\def\gsim{{~\raise.15em\hbox{$>$}\kern-.85em
          \lower.35em\hbox{$\sim$}~}\xspace}
\def\lsim{{~\raise.15em\hbox{$<$}\kern-.85em
          \lower.35em\hbox{$\sim$}~}\xspace}

\def\PDF {PDF\xspace}

\def\sPlot{\mbox{\em sPlot}}
\def\sWeight{\mbox{\em sWeight}}

\def\evtgen     {\mbox{\textsc{EvtGen}}\xspace}

\def\tell1  {TELL1\xspace}
\def\ukl1   {UKL1\xspace}

\usepackage{cite} 
\usepackage{mciteplus}

\begin{document}

\renewcommand{\thefootnote}{\fnsymbol{footnote}}
\setcounter{footnote}{1}


\begin{titlepage}
\pagenumbering{roman}

\vspace*{-1.5cm}
\centerline{\large EUROPEAN ORGANIZATION FOR NUCLEAR RESEARCH (CERN)}
\vspace*{1.5cm}
\hspace*{-0.5cm}
\begin{tabular*}{\linewidth}{lc@{\extracolsep{\fill}}r}
\ifthenelse{\boolean{pdflatex}}
{\vspace*{-2.7cm}\mbox{\!\!\!\includegraphics[width=.14\textwidth]{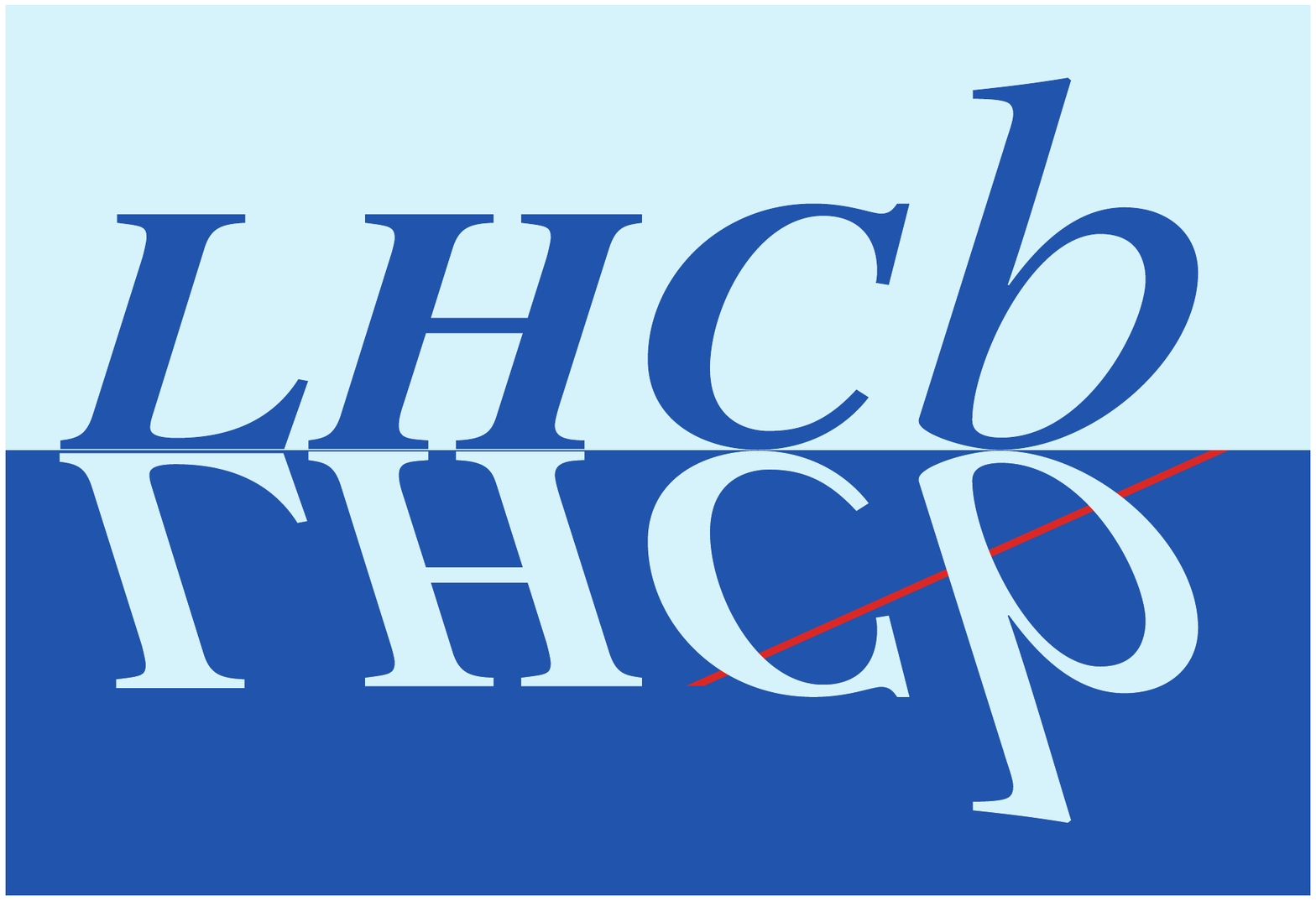}} & &}%
{\vspace*{-1.2cm}\mbox{\!\!\!\includegraphics[width=.12\textwidth]{lhcb-logo.eps}} & &}%
\\
 & & CERN-PH-EP-2013-017 \\  
 & & LHCb-PAPER-2013-001 \\  
 & & 25 February 2013 \\ 
 & & \\
\end{tabular*}

\vspace*{4.0cm}

{\bf\boldmath\huge
\begin{center}
  Determination of the $X(3872)$ meson quantum numbers
\end{center}
}

\vspace*{2.0cm}

\begin{center}
The LHCb collaboration\footnote{Authors are listed on the following pages.}
\end{center}

\vspace{\fill}

\begin{abstract}
  \noindent
The quantum numbers of the $X(3872)$ meson are determined to be $J^{PC} = 1^{++}$ 
based on angular correlations in 
$B^+\to X(3872) K^+$ decays, where $X(3872)\to \pi^+\pi^- J/\psi$ and $J/\psi \to\mu^+\mu^-$.
The data correspond to 1.0 fb$^{-1}$ of $pp$ collisions collected by the LHCb detector. 
The only alternative assignment allowed by previous measurements, $J^{PC}=2^{-+}$,
is rejected with a confidence level 
equivalent to more than eight Gaussian standard deviations
using the likelihood-ratio test in the full angular phase space. 
This result favors exotic explanations of the $X(3872)$ state.

\end{abstract}

\vspace*{2.0cm}

\begin{center}
  Submitted to Physical Review Letters
\end{center}

\vspace{\fill}

{\footnotesize 
\centerline{\copyright~CERN on behalf of the \lhcb collaboration, license \href{http://creativecommons.org/licenses/by/3.0/
}{CC-BY-3.0}.}}
\vspace*{2mm}

\end{titlepage}


\newpage
\setcounter{page}{2}
\mbox{~}
\newpage

\centerline{\large\bf LHCb collaboration}
\begin{flushleft}
\small
R.~Aaij$^{40}$, 
C.~Abellan~Beteta$^{35,n}$, 
B.~Adeva$^{36}$, 
M.~Adinolfi$^{45}$, 
C.~Adrover$^{6}$, 
A.~Affolder$^{51}$, 
Z.~Ajaltouni$^{5}$, 
J.~Albrecht$^{9}$, 
F.~Alessio$^{37}$, 
M.~Alexander$^{50}$
% ...

{\footnotesize \it
$ ^{1}$Centro Brasileiro de Pesquisas F\'{i}sicas (CBPF), Rio de Janeiro, Brazil\\
% ...
$ ^{5}$Clermont Universit\'{e}, Universit\'{e} Blaise Pascal, CNRS/IN2P3, LPC, Clermont-Ferrand, France\\
$ ^{6}$CPPM, Aix-Marseille Universit\'{e}, CNRS/IN2P3, Marseille, France\\
% ...
$ ^{9}$Fakult\"{a}t Physik, Technische Universit\"{a}t Dortmund, Dortmund, Germany\\
% ...
$ ^{35}$Universitat de Barcelona, Barcelona, Spain\\
$ ^{36}$Universidad de Santiago de Compostela, Santiago de Compostela, Spain\\
$ ^{37}$European Organization for Nuclear Research (CERN), Geneva, Switzerland\\
% ...
$ ^{40}$Nikhef National Institute for Subatomic Physics, Amsterdam, The Netherlands\\
% ...
$ ^{45}$H.H. Wills Physics Laboratory, University of Bristol, Bristol, United Kingdom\\
% ...
$ ^{50}$School of Physics and Astronomy, University of Glasgow, Glasgow, United Kingdom\\
$ ^{51}$Oliver Lodge Laboratory, University of Liverpool, Liverpool, United Kingdom\\
% ...
$ ^{n}$LIFAELS, La Salle, Universitat Ramon Llull, Barcelona, Spain\\
% ...
}
\end{flushleft}

\cleardoublepage


\renewcommand{\thefootnote}{\arabic{footnote}}
\setcounter{footnote}{0}



\pagestyle{plain} 
\setcounter{page}{1}
\pagenumbering{arabic}


%

\newboolean{prl}
\setboolean{prl}{false} 

\newlength{\figsize}
\setlength{\figsize}{0.8\hsize}
\def\bpsi2skp{\bar{B}^0\to\psi(2S)K^-\pi^+}
\def\bjpsikp{\bar{B}^0\to\jpsi K^-\pi^+}
\def\bjpsiks{\bar{B}^0\to\jpsi K^*}
\def\bpsiks{\bar{B}^0\to\psi K^*}
\def\bchic1kp{\bar{B}^0\to\chi_{c1} K^-\pi^+}
\def\bcjppp{B_c^+\to\jpsi\pi^+\pi^-\pi^+}
\def\bjkkk{B^+\to\jpsi K^+K^-K^+}
\def\bjkpp{B^+\to\jpsi K^+\pi^+\pi^-}
\def\Kstar{K^*}
\def\BR{{\cal B}}
\def\DLL{DLL}
\def\PDF{{\cal P}}
\def\M{{\cal M}}
\def\NDOF{\hbox{\rm ndf}}
\def\cospsi{\cos\theta_{\psi}}
\def\cospsiz{\cos\theta_{\psi}^Z}
\def\cosks{\cos\theta_{K^*}}
\def\cosz{\cos\theta_{Z}}
\def\mkp{m_{K\pi}}
\def\mkpi{\mkp}
\def\mpsi2sp{m_{\psi(2S)\pi}}
\def\mpsip{m_{\psi\pi}}
\def\mpsipi{\mpsip}
\def\Sum{\sum}
\def\Int{\int}
\def\Frac{\frac}
\def\keoo{K^*_0(800)}
\def\kent{K^*(892)}
\def\xff{X(3872)}
\def\bpsi2sk{B^+\to\psi(2S) K^+}
\def\bxk{B^+\to\xff K^+}
\def\bjks{B^+\to\jpsi K^{*+}}
\def\kone{K_1(1270)^+}
\def\bjkone{B^+\to\jpsi\kone}
\def\xppj{\xff\to\pi^+\pi^-\jpsi}
\def\xrj{\xff\to\rho(770)\jpsi}
\def\rpp{\rho(770)\to\pi^+\pi^-}
\def\pipi{\pi\pi}
\def\jll{\jpsi\to \ell^+\ell^-}
\def\jmm{\jpsi\to \mu^+\mu^-}
\def\jpc{J^{PC}}
\def\DM{\Delta M}
\def\mjjp{\DM} 
\def\mppj{\DM} 
\def\mjppj{\DM} 
\def\jppk{\jpsi\pi^+\pi^-K^+}
\def\jkpp{\jpsi\pi^+\pi^-K^+}
\def\pp{\pi\pi}
\def\mpp{M(\pi^+\pi^-)}
\def\Q{M(\jpsi\pi^+\pi^-)-M(\jpsi)-\mpp}
\def\L{{\cal L}}
\def\sWeights{{\mbox{\em sWeights}}}

\noindent
It has been almost ten years since the narrow $\xff$ state 
was discovered in $B^+$ decays
by the Belle experiment 
\cite{Choi:2003ue}.\footnote{The inclusion of charge-conjugate states is implied in this Letter.}
Subsequently, its existence has been confirmed by several
other experiments \cite{CDFPhysRevLett.93.072001,D0Abazov:2004kp,BaBarPhysRevD.71.071103}.
Recently, its production has been studied at the LHC \cite{Aaij:2011sn,Chatrchyan:2013cld}.
However, the nature of this state remains unclear. 
Among the open possibilities are 
conventional charmonium and exotic states such as 
$D^{*0}\bar{D}^0$ molecules \cite{Tornqvist:2004qy}, 
tetra-quarks \cite{Maiani:2004vq} or their mixtures \cite{Hanhart:2011jz}.
Determination of the quantum numbers, 
total angular momentum $J$, parity $P$, 
and charge-conjugation $C$, 
is important to shed light on this ambiguity.
The $C$-parity of the state is positive
since the $\xff\to\gamma\jpsi$ decay has been observed \cite{Aubert:2006aj,Bhardwaj:2011dj}.

The CDF experiment analyzed three-dimensional (3D) 
angular correlations in a relatively high-background
sample of $2292\pm113$ 
inclusively-reconstructed $\xppj$, $\jmm$ decays, 
dominated by prompt production in $p\bar p$ collisions.
The unknown polarization of the $\xff$ mesons limited 
the sensitivity of the measurement of $\jpc$ \cite{Abulencia:2006ma}.
A $\chi^2$ fit of $\jpc$ hypotheses to the binned 3D distribution
of the $\jpsi$ and $\pipi$ helicity angles 
($\theta_{\jpsi}$, $\theta_{\pipi}$) \cite{Jacob:1959at,Richman:1984gh,PhysRevD.57.431},
and the angle between their decay planes 
($\Delta\phi_{\jpsi,\pipi}=\phi_{\jpsi}-\phi_{\pipi}$),
excluded all spin-parity assignments except for $1^{++}$ or $2^{-+}$.
The Belle collaboration observed 
$173\pm16$ $B\to\xff K$ ($K=K^{\pm}$ or $K^0_S$), $\xppj$, $\jll$ decays \cite{Choi:2011fc}.
The reconstruction of the full decay chain resulted in a small background and polarized $\xff$ mesons,
making their helicity angle ($\theta_X$) and orientation of their decay plane ($\phi_X$) sensitive 
to $\jpc$ as well. 
By studying one-dimensional distributions in three different angles,
they concluded that their data were equally well described by the $1^{++}$ and $2^{-+}$ hypotheses. 
The BaBar experiment observed $34\pm7$ $\xff\to\omega\jpsi$, 
$\omega\to\pi^+\pi^-\pi^0$ events \cite{delAmoSanchez:2010jr}.
The observed $\pi^+\pi^-\pi^0$ mass distribution 
favored the $2^{-+}$ hypothesis, which had a confidence level ({\rm CL}) of $68\%$, 
over the $1^{++}$ hypothesis, but the latter was not ruled out ({\rm CL}~$=7\%$).

In this Letter, we report the first analysis of the complete five-dimensional angular correlations 
of the $B^+\to\xff K^+$, $\xppj$, $\jmm$ decay chain 
using $\sqrt{s}=7$~TeV $pp$ collision data corresponding to $1.0$~fb$^{-1}$  
collected in 2011 by the LHCb experiment.
The \lhcb detector~\cite{Alves:2008zz} is a single-arm forward
spectrometer covering the \mbox{pseudorapidity} range $2<\eta <5$,
designed for the study of particles containing \bquark or \cquark
quarks. The detector includes a high precision tracking system
consisting of a silicon-strip vertex detector surrounding the $pp$
interaction region, a large-area silicon-strip detector located
upstream of a dipole magnet with a bending power of about
$4{\rm\,Tm}$, and three stations of silicon-strip detectors and straw
drift tubes placed downstream. The combined tracking system has 
momentum resolution $\Delta p/p$ that varies from 0.4\% at 5\gev to
0.6\% at 100\gev, and impact parameter (IP) resolution of 20\mum for
tracks with high transverse momentum ($p_{\rm T}$).\footnote{We 
use mass and momentum units in which $c=1$.}  
Charged hadrons are identified
using two ring-imaging Cherenkov detectors. Photon, electron and
hadron candidates are identified by a calorimeter system consisting of
scintillating-pad and preshower detectors, an electromagnetic
calorimeter and a hadronic calorimeter. Muons are identified by a
system composed of alternating layers of iron and multiwire
proportional chambers. The trigger~\cite{Aaij:2012me} consists of a
hardware stage, based on information from the calorimeter and muon
systems, followed by a software stage which applies a full event
reconstruction.
 
In the offline analysis 
$\jpsi\to\mu^+\mu^-$ candidates are selected with the following criteria: 
$p_{\rm T}(\mu)>0.9$~GeV,
$p_{\rm T}(\jpsi)>1.5$~GeV,   
$\chi^2$ per degree of freedom for the two muons to form a common vertex, $\chi^2_{\rm vtx}(\mu^+\mu^-)/\NDOF<9$,
and a mass consistent with the $\jpsi$ meson. 
The separation of the $\jpsi$ decay vertex from the nearest primary vertex (PV) must be at least 
three standard deviations.    
Combinations of $K^+\pi^-\pi^+$ candidates that are consistent with originating from a common vertex with 
$\chi^2_{\rm vtx}(K^+\pi^-\pi^+)/\NDOF<9$, 
with each charged hadron ($h$) separated from all PVs ($\chi^2_{\rm IP}(h)>9$)
and having $p_{\rm T}(h)>0.25$~GeV, are selected.
The quantity $\chi^2_{\rm IP}(h)$ is defined as the
difference between the \chisq of the PV reconstructed with and
without the considered particle. 
Kaon and pion candidates are required to satisfy $\ln[\L(K)/\L(\pi)]>0$ and $<5$, respectively,
where $\L$ is the particle identification likelihood \cite{arXiv:1211-6759}.
If both same-sign hadrons in this combination meet the kaon requirement, only
the particle with higher $p_{\rm T}$ is considered a kaon candidate. 
We combine $\jpsi$ candidates with $K^+\pi^-\pi^+$ candidates to form 
$B^+$ candidates, which must satisfy $\chi^2_{\rm vtx}(\jpsi K^+\pi^-\pi^+)/\NDOF<9$, $p_{\rm T}(B^+)>2$ GeV 
and have decay time greater than $0.25$~ps. The $\jpsi K^+\pi^-\pi^+$ mass is calculated 
using the known $\jpsi$ mass and 
the $B$ vertex as constraints.

Four discriminating variables ($x_i$) are
used in a likelihood ratio to improve the background suppression:
the minimal $\chi^2_{\rm IP}(h)$,
$\chi^2_{\rm vtx}(\jpsi  K^+\pi^+\pi^-)/\NDOF$,
$\chi^2_{\rm IP}(B^+)$,
and the cosine of the largest opening angle between the $J/\psi$ and 
the charged-hadron transverse momenta. 
The latter peaks at positive values for the signal as the $B^+$ meson has a high transverse momentum.
Background events in which particles are combined 
from two different $B$ decays peak at negative values,
whilst those due to random combinations of particles are more uniformly distributed.  
The four 1D signal probability density functions (PDFs), $\PDF_{\rm sig}(x_i)$, are obtained from a simulated 
sample of $B^+\to\psi(2S)K^+$, $\psi(2S)\to\pi^+\pi^-\jpsi$ decays,
which are kinematically similar to the signal decays.
The data sample of $B^+\to\psi(2S)K^+$ events 
is used as a control sample for $\PDF_{\rm sig}(x_i)$
and for systematic studies in the angular analysis.
The background PDFs, $\PDF_{\rm bkg}(x_i)$, are obtained from the data 
in the $B^+$ mass sidebands (4.85--5.10 and 5.45--6.50 GeV).
We require $-2 \sum_{i=1}^4 \ln[\PDF_{\rm sig}(x_i)/\PDF_{\rm bkg}(x_i)]<1.0$,
which preserves about 94\%\ of the $\xff$ signal events.

\begin{figure}[htbp]
  \begin{center}
  \ifthenelse{\boolean{pdflatex}}{ 
        \includegraphics*[width=\figsize]{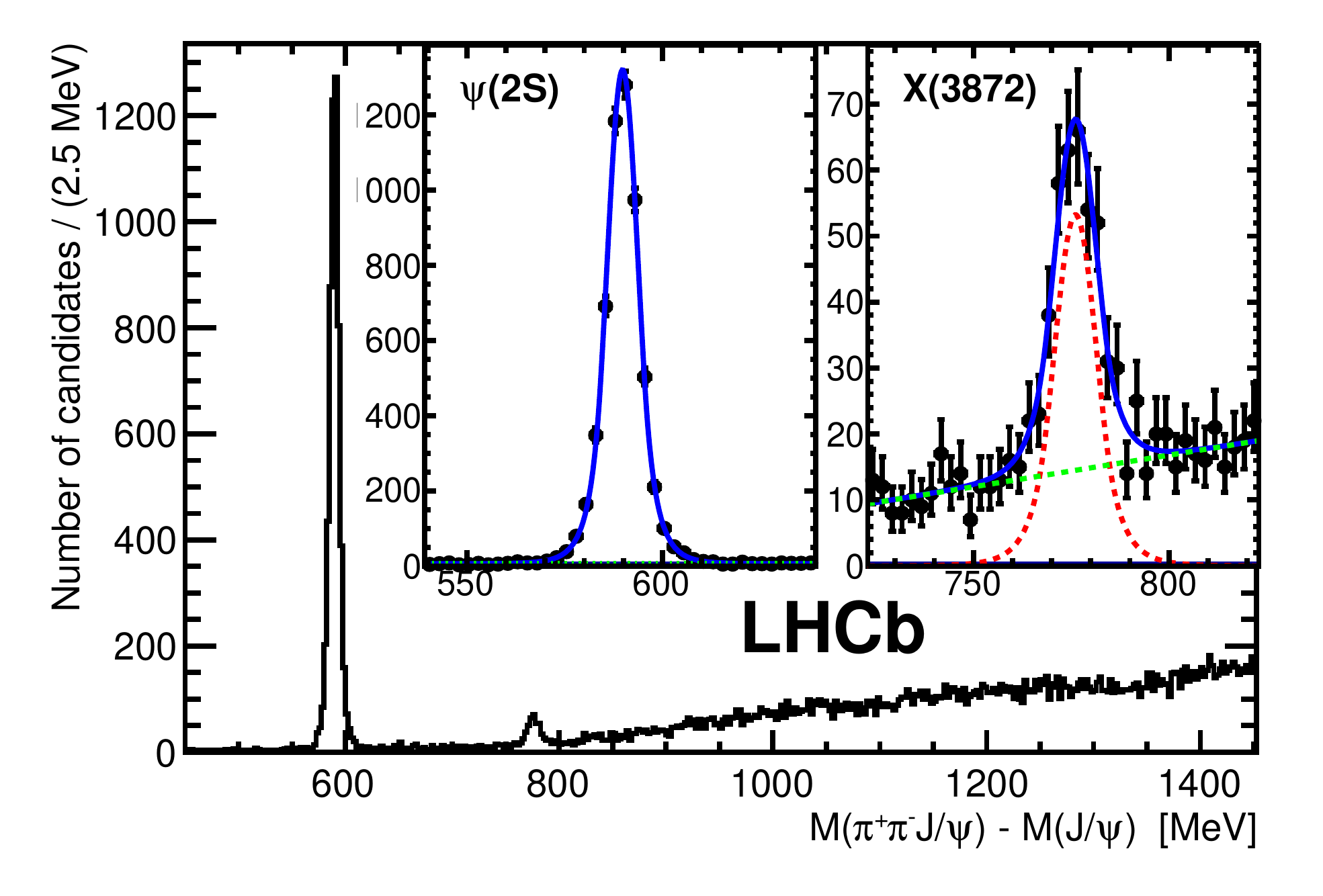}
   }{
        \includegraphics*[width=\figsize]{jppj.eps}
   } 
  \end{center}
  \vskip-1.1cm\caption{\small 
    Distribution of $\mppj$ for $\bjkpp$ candidates.
    The fits of the $\psi(2S)$ and $\xff$ signals are displayed.
    The solid blue, dashed red, and dotted green lines represent 
    the total fit, signal component, and background component, 
    respectively.
  \label{fig:jppj}
  }
\end{figure}

About $38\,000$ candidates are selected  
in a $\pm2\sigma$ mass range 
around the $B^+$ peak 
in the $M(\jkpp)$ distribution, 
with a signal purity of 89\%. 
The $\Delta M=M(\pi^+\pi^-\jpsi)-M(\jpsi)$ distribution is shown in Fig.~\ref{fig:jppj}.
Fits to the $\psi(2S)$ and $\xff$ signals are shown in the insets.
A Crystal Ball function \cite{Skwarnicki:1986xj} with symmetric tails is used for the signal shapes.
The background is assumed to be linear.
The $\psi(2S)$ fit is performed in the 539.2--639.2 MeV range leaving all parameters free to vary.  
It yields $5642\pm76$ signal ($230\pm21$ background) candidates with a $\Delta M$ resolution of   
$\sigma_{\Delta M}=3.99\pm0.05$ MeV,
corresponding to a signal purity of $99.2\%$ within 
a $\pm2.5\sigma_{\Delta M}$ region.   
When fitting in the 723--823 MeV range, 
the signal tail parameters are fixed to the values obtained in the $\psi(2S)$ fit, 
which also describe well the simulated $\xff$ signal distribution.
The fit yields  $313\pm26$ $\bxk$ ($568\pm31$ background) candidates with a resolution of   
$5.5\pm0.5$ MeV,   
corresponding to a signal purity of $68\%$ within 
a $\pm2.5\sigma_{\Delta M}$ region.   
The dominant source of background 
is from $B^+\to \jpsi K_1(1270)^+$, $K_1(1270)^+\to K^+\pi^+\pi^-$ decays
as found by studying the $K^+\pi^+\pi^-$ mass distribution.  

The angular correlations in the $B^+$ decay  
carry information about the $\xff$ quantum numbers.
To discriminate between the $1^{++}$ and $2^{-+}$ assignments we
use the likelihood-ratio test, 
which in general provides the most powerful test 
between two hypotheses \cite{james2006statistical}. 
The PDF for each $\jpc$ hypothesis, $J_X$,  is defined
in the 5D angular space $\Omega\equiv$
$(\cos\theta_X,\cos\theta_{\pipi},\Delta\phi_{X,\pipi},\cos\theta_{\jpsi},\Delta\phi_{X,\jpsi})$
by the normalized product of
the expected decay matrix element ($\M$) squared
and of the reconstruction efficiency ($\epsilon$), 
$\PDF(\Omega|J_X)=|\M(\Omega|J_X)|^2\,\epsilon(\Omega)/I(J_X)$, where
$I(J_X)=\int|\M(\Omega|J_X)|^2\,\epsilon(\Omega){\it d}\Omega$.
The efficiency is averaged over the $\pi^+\pi^-$ mass ($M(\pipi)$) using  
a simulation \cite{Sjostrand:2006za,LHCb-PROC-2010-056,Lange:2001uf,Allison:2006ve,*Agostinelli:2002hh,LHCb-PROC-2011-006}
that assumes 
the $\xff\to\rho(770)\jpsi$, $\rho(770)\to\pi^+\pi^-$ decay 
\cite{Choi:2011fc,Abulencia:2005zc,Chatrchyan:2013cld}.
The observed $M(\pipi)$ distribution is in good agreement with this simulation.
The lineshape of the $\rho(770)$ resonance can change slightly depending on
the spin hypothesis.
The effect on $\epsilon(\Omega)$ is found to be very small and is neglected. 
We follow the approach adopted  
in Ref.~\cite{Abulencia:2006ma} 
to predict the matrix elements. 
The angular correlations are obtained using the helicity formalism,

\ifthenelse{\boolean{prl}}{ 
\begin{eqnarray*}
 |\, \M(\Omega|J_X)\,|^2 & = & \sum_{\Delta\lambda_{\mu}=-1,+1}  \left| \phantom{A_{\lambda_\psi}}\hskip-0.4cm \right.
 \sum_{\lambda_{\jpsi},\lambda_{\pp}=-1,0,+1}  \notag\\
A_{\lambda_{\jpsi},\lambda_{\pp}} &  
 \times & D^{J_X}_{0\,,\,\lambda_{\jpsi}-\lambda_{\pp}}(\phi_X,\theta_X,-\phi_X) \times \notag\\
& & D^{1}_{\lambda_{\pp}\,,\,0}(\phi_{\pp},\theta_{\pp},-\phi_{\pp}) \times \notag\\
& & 
D^{1}_{\lambda_{\jpsi}\,,\,\Delta\lambda_{\mu}}(\phi_{\jpsi},\theta_{\jpsi},-\phi_{\jpsi})
\left. \phantom{A_{\lambda_\psi}}\hskip-0.4cm \right|^2, \notag\\
\end{eqnarray*}
}{
\begin{eqnarray*}  |\, \M(\Omega|J_X)\,|^2 & = & \sum_{\Delta\lambda_{\mu}=-1,+1}  \left| \phantom{A_{\lambda_\psi}}\hskip-0.4cm \right.  
\sum_{\lambda_{\jpsi},\lambda_{\pp}=-1,0,+1}  A_{\lambda_{\jpsi},\lambda_{\pp}} \times D^{J_X}_{0\,,\,\lambda_{\jpsi}-\lambda_{\pp}}(\phi_X,\theta_X,-\phi_X) \times \notag\\
 & & \hskip2.4cm D^{1}_{\lambda_{\pp}\,,\,0}(\phi_{\pp},\theta_{\pp},-\phi_{\pp}) \times  D^{1}_{\lambda_{\jpsi}\,,\,\Delta\lambda_{\mu}}(\phi_{\jpsi},\theta_{\jpsi},-\phi_{\jpsi})  
\left. \phantom{A_{\lambda_\psi}}\hskip-0.4cm \right|^2 , \notag\\  
\end{eqnarray*}
}

where $\lambda$ are particle helicities and $D^J_{\lambda_1\,,\,\lambda_2}$ 
are Wigner functions \cite{Jacob:1959at,Richman:1984gh,PhysRevD.57.431}.
The helicity couplings, $A_{\lambda_{\jpsi},\lambda_{\pp}}$,
are expressed in terms of the $LS$ couplings \cite{IEKP-KA/2008-16,CERN-THESIS-2012-003}, 
$B_{LS}$, where $L$ is the orbital angular momentum between the $\pipi$ system and 
the $\jpsi$ meson, and 
$S$ is the sum of their spins.
Since the energy release in the $\xff\to\rho(770)\jpsi$
decay is small, the lowest value of $L$ 
is expected to dominate, especially because the
next-to-minimal value is not allowed by parity conservation.
The lowest value for the $1^{++}$ hypothesis is $L=0$, which implies $S=1$. 
With only one $LS$ amplitude present, the angular distribution 
is completely determined without free parameters.
For the $2^{-+}$ hypothesis the lowest value is $L=1$, which implies $S=1$ or $2$. 
As both $LS$ combinations are possible, the $2^{-+}$ hypothesis
implies two parameters, which are chosen to be the real and imaginary
parts of $\alpha\equiv B_{11}/(B_{11}+B_{12})$.
Since they are related to strong dynamics, they are difficult to predict 
theoretically and are treated as nuisance parameters.

We define a test statistic $t=-2\ln[\L(2^{-+})/\L(1^{++})]$, 
where the $\L(2^{-+})$ likelihood is maximized with respect to $\alpha$.   
The efficiency $\epsilon(\Omega)$ is not
determined on an event-by-event basis, 
since it cancels in the likelihood ratio except for the normalization integrals.
A large sample of simulated events, with uniform angular distributions,
passed through a full simulation of the detection and the data selection process,
is used to carry out the integration, $I(J_X)\propto \sum_{i=1}^{N_{\rm MC}} |\M(\Omega_i|J_X)|^2$, 
where $N_{\rm MC}$ is the number of reconstructed simulated events.
The background in the data is subtracted in the log-likelihoods
using the \sPlot\ technique \cite{2005NIMPA.555..356P} 
by assigning to each candidate in the fitted $\Delta M$ range
an event weight (\sWeight), $w_i$, based on its $\Delta M$ value,
$-2\ln \L(J_X)= - s_w\,2\sum_{i=1}^{N_{\rm data}} w_i\, \ln \PDF(\Omega_i|J_X)$.
Here, $s_w$ is a constant scaling factor, 
$s_w = \sum_{i=1}^{N_{\rm data}} w_i/\sum_{i=1}^{N_{\rm data}} {w_i}^2$,
which accounts for statistical fluctuations in the background subtraction. 
Positive (negative) values of the test statistic for the data,
$t_{\rm data}$, favor the $1^{++}$ ($2^{-+}$) hypothesis.
The analysis procedure has been extensively tested on simulated samples
for the $1^{++}$ and $2^{-+}$ hypotheses with different values of $\alpha$,  
generated using 
the \evtgen\ package \cite{Lange:2001uf}.

\begin{figure}[htbp]
  \begin{center}
  \ifthenelse{\boolean{pdflatex}}{
    \includegraphics*[width=\figsize]{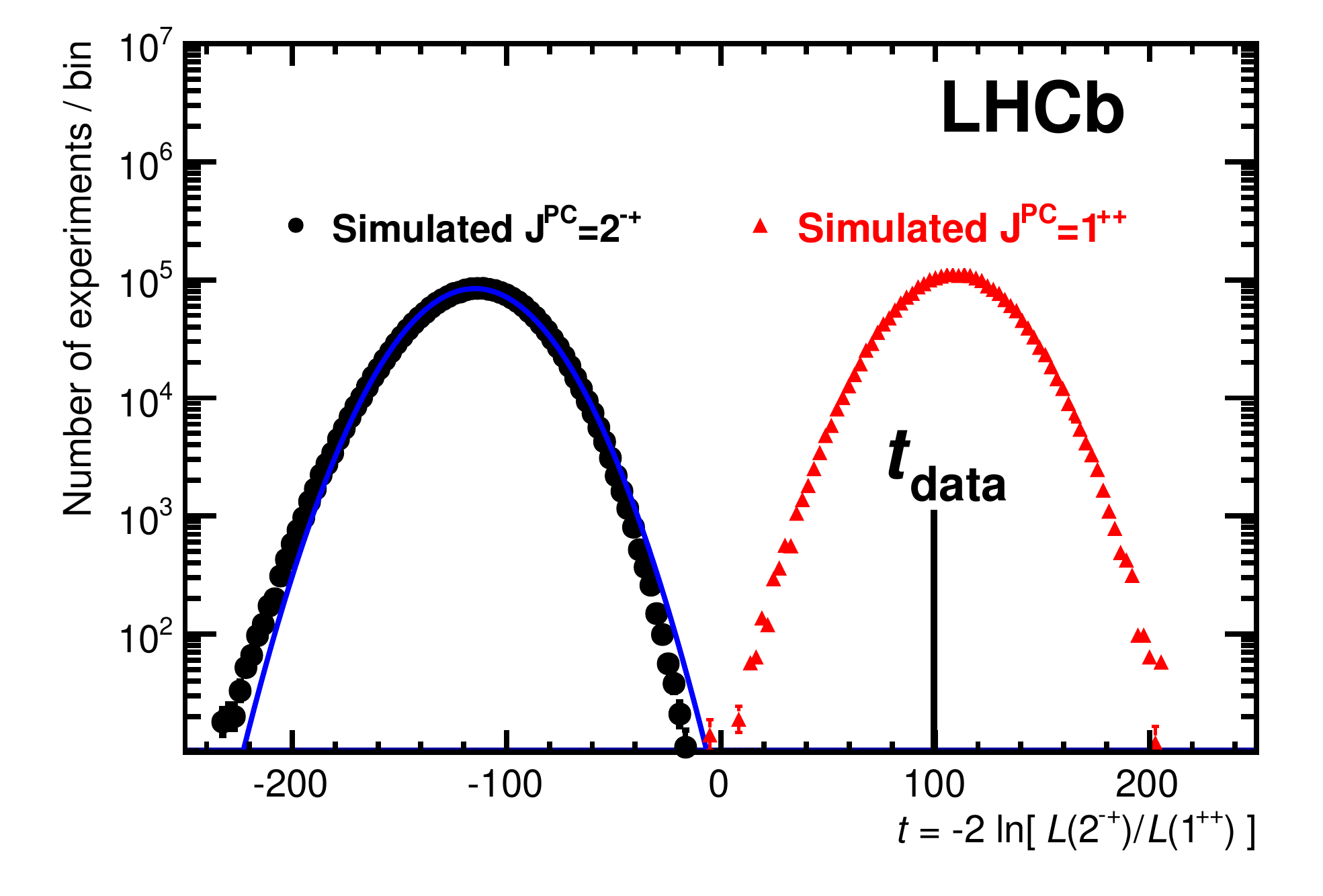} \quad 
   }{
    \includegraphics*[width=\figsize]{dchi2L_log.eps}
   } 
  \end{center}
  \vskip-1.1cm\caption{\small
    Distribution of the test statistic $t$ 
    for the simulated experiments 
    with $\jpc=2^{-+}$ and $\alpha=\hat\alpha$ 
    (black circles on the left)
    and with $\jpc=1^{++}$ (red triangles on the right).
    A Gaussian fit to the $2^{-+}$ distribution 
    is overlaid (blue solid line).
    The value of the test statistic for the data, $t_{\rm data}$,
    is shown by the solid vertical line. 
  \label{fig:toys}
  }
\end{figure}

\begin{figure}[htbp]
  \begin{center}
  \ifthenelse{\boolean{pdflatex}}{
    \includegraphics*[width=\figsize]{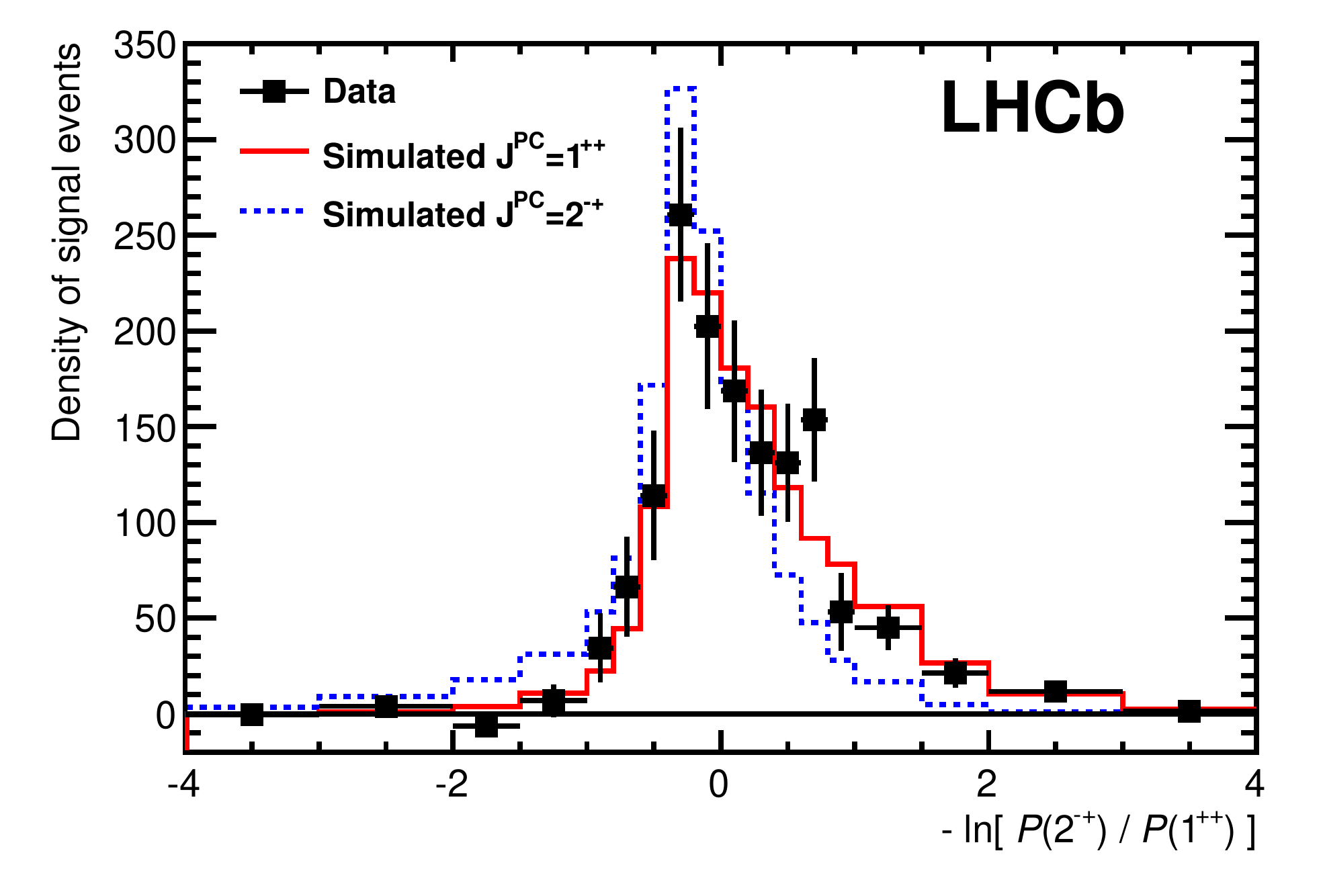}  
   }{
    \includegraphics*[width=\figsize]{llr.eps}
   } 
  \end{center}
  \vskip-1.1cm\caption{\small
    Distribution of $-\ln[{\PDF(\Omega_i|2^{-+},\hat\alpha)}/{\PDF(\Omega_i|1^{++})}]$
    for the data (points with error bars)
    compared to the distributions for the simulated experiments 
    with $\jpc=1^{++}$ (red solid histogram) 
    and with $\jpc=2^{-+}$, $\alpha=\hat\alpha$ (blue dashed histogram)
    after the background subtraction using \sWeights.
    The simulated distributions are normalized to the number of signal candidates
    observed in the data.
    Bin contents and uncertainties are divided by bin width because of unequal bin sizes.      
  \label{fig:llr}
  }
\end{figure}

The value of $\alpha$ that minimizes $-2\ln\L(J_X=2^{-+},\alpha)$ in the data is 
$\hat\alpha=(0.671\pm0.046,0.280\pm0.046)$. This is compatible with the value reported by Belle, 
$(0.64,0.27)$ \cite{Choi:2011fc}.
The value of the test statistic observed in the data is 
$t_{\rm data}=+99$, thus favoring the $1^{++}$ hypothesis. 
Furthermore, $\hat\alpha$ is 
consistent with the value of $\alpha$ obtained from fitting a large background-free 
sample of simulated $1^{++}$ events, $(0.650\pm0.011,0.294\pm0.012)$.
The value of $t_{\rm data}$ is compared with the distribution 
of $t$ in the simulated experiments
to determine a $p$-value for the $2^{-+}$ hypothesis 
via the fraction of simulated experiments yielding a value of $t>t_{\rm data}$. 
We simulate 2 million experiments with the value of $\alpha$, 
and the number of signal and background events, as observed in the data.
The background is assumed to be saturated by the $\bjkone$ decay, which provides
a good description of its angular correlations. 
None of the values of $t$ from the simulated experiments   
even approach $t_{\rm data}$, indicating a $p$-value smaller than
$1/(2\times10^6)$, which corresponds to 
a rejection of the $2^{-+}$ hypothesis with greater than $5\sigma$ significance. 
As shown in Fig.~\ref{fig:toys}, 
the distribution of $t$ is
reasonably well approximated by a Gaussian function.
Based on the mean and r.m.s.\ spread of the $t$ distribution for 
the $2^{-+}$ experiments, this hypothesis is rejected with a significance of $8.4\sigma$. 
The deviations of the $t$ distribution from
the Gaussian function  
suggest this is a plausible estimate.
Using phase-space $\bjkpp$ decays as a model for the background 
events, we obtain a consistent result. 
The value of $t_{\rm data}$ falls into the region where 
the probability density for the $1^{++}$ simulated experiments is high. 
Integrating the $1^{++}$ distribution from $-\infty$ to $t_{\rm data}$
gives ${\rm CL}~(1^{++})=34\%$. 
We also compare the binned distribution of 
single-event log-likelihood-ratios with \sWeights\ applied, 
$\ln[\PDF(\Omega_i|2^{-+},\hat\alpha)/\PDF(\Omega_i|1^{++})]$,
between the data and the simulations.
The shape of this distribution in data is consistent with the 
$1^{++}$ simulations 
and inconsistent with the $2^{-+}$ simulations, as illustrated in Fig.~\ref{fig:llr}.

We vary the data selection criteria to probe for possible
biases from the background subtraction and the efficiency
corrections. The nominal selection does not bias the $M(\pipi)$
distribution. 
By requiring $Q=M(\jpsi\pipi)-M(\jpsi)-M(\pipi)<0.1$ GeV, we reduce 
the background level by a factor of four, 
while losing only $21\%$ of the signal. 
The significance of the $2^{-+}$ rejection changes very little,
in agreement with the simulations. 
By tightening the requirements on the $p_{\rm T}$ of $\pi$, $K$ and $\mu$ candidates,
we decrease the signal efficiency 
by about 50\%\
with similar reduction
in the background level.
In all cases, the significance of the $2^{-+}$ rejection 
is reduced by a factor consistent with the 
simulations.

In the analysis we use simulations to calculate the $I(J_X)$ integrals.
In an alternative approach to the efficiency estimates, 
we use the $\bpsi2sk$ events observed
in the data weighted 
by the inverse of 
$1^{--}$ matrix element squared.  
We obtain a value of $t_{\rm data}$ that corresponds to
$8.2\sigma$ rejection of the $2^{-+}$ hypothesis. 

\begin{figure}[htbp]
  \begin{center}
  \ifthenelse{\boolean{pdflatex}}{
        \includegraphics*[width=0.8\figsize]{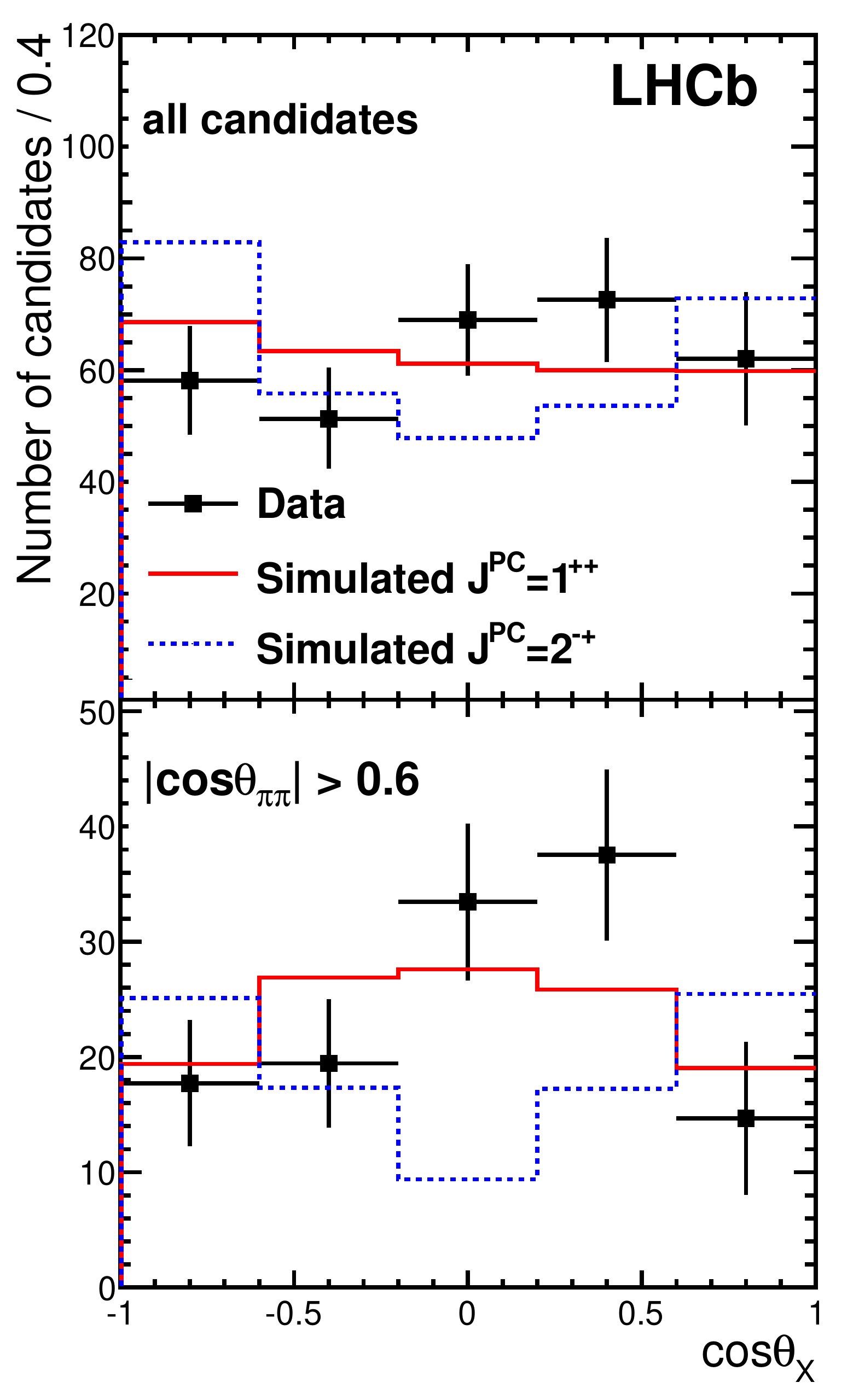}
   }{
        \includegraphics*[width=0.8\figsize]{cosxcosr_new.eps}
  }
  \end{center}
  \vskip-1.1cm\caption{\small
    Background-subtracted 
    distribution of $\cos\theta_X$ for (top) all candidates and for (bottom) 
    candidates with $|\cos\theta_{\pp}|>0.6$ 
    for the data (points with error bars)
    compared to the expected distributions for the 
    $\jpc=1^{++}$ (red solid histogram) 
    and $\jpc=2^{-+}$ and $\alpha=\hat\alpha$ hypotheses (blue dashed histogram).
    The simulated distributions are normalized 
    to the number of signal candidates observed in the data
    across the full phase space.
  \label{fig:cosxcosr}
  }
\end{figure}

As an additional goodness-of-fit test for the $1^{++}$ hypothesis, we project 
the data onto five 1D and ten 2D binned distributions in all five angles and their
combinations. 
They are all consistent with the distributions expected for the $1^{++}$ hypothesis.
Some of them are inconsistent with the distributions expected for 
the ($2^{-+}$, $\hat\alpha$) hypothesis. 
The most significant inconsistency is observed for the 2D projections 
onto $\cos\theta_X$ vs.\ $\cos\theta_{\pipi}$. 
The separation between the $1^{++}$ and $2^{-+}$ hypotheses increases when using
correlations between these two angles, as illustrated in Fig.~\ref{fig:cosxcosr}.

In summary, we unambiguously establish
that the values of total angular momentum, 
parity and charge-conjugation eigenvalues of the $X(3872)$ state
are $1^{++}$. 
This is achieved through the first analysis of the full five-dimensional 
angular correlations between final state particles
in $B^+\to\xff K^+$, $\xppj$, $\jmm$ decays using the likelihood-ratio test. 
The $2^{-+}$ hypothesis is excluded with a significance of 
more than eight Gaussian standard deviations.
This result rules out the explanation of the $\xff$ meson as a conventional 
$\eta_{c2}(1^1D_2)$ state. Among the remaining possibilities are 
the $\chi_{c1}(2^3P_1)$ charmonium, disfavored by the value of  
the $\xff$ mass \cite{Skwarnicki:2003wn},
and unconventional explanations such 
as a $D^{*0}\bar{D}^0$ molecule \cite{Tornqvist:2004qy},
tetraquark state \cite{Maiani:2004vq} or
charmonium-molecule mixture \cite{Hanhart:2011jz}.


\section*{Acknowledgements}

\noindent We express our gratitude to our colleagues in the CERN
accelerator departments for the excellent performance of the LHC. We
thank the technical and administrative staff at the LHCb
institutes. We acknowledge support from CERN and from the national
agencies: CAPES, CNPq, FAPERJ and FINEP (Brazil); NSFC (China);
CNRS/IN2P3 and Region Auvergne (France); BMBF, DFG, HGF and MPG
(Germany); SFI (Ireland); INFN (Italy); FOM and NWO (The Netherlands);
SCSR (Poland); ANCS/IFA (Romania); MinES, Rosatom, RFBR and NRC
``Kurchatov Institute'' (Russia); MinECo, XuntaGal and GENCAT (Spain);
SNSF and SER (Switzerland); NAS Ukraine (Ukraine); STFC (United
Kingdom); NSF (USA). We also acknowledge the support received from the
ERC under FP7. The Tier1 computing centres are supported by IN2P3
(France), KIT and BMBF (Germany), INFN (Italy), NWO and SURF (The
Netherlands), PIC (Spain), GridPP (United Kingdom). We are thankful
for the computing resources put at our disposal by Yandex LLC
(Russia), as well as to the communities behind the multiple open
source software packages that we depend on.

\addcontentsline{toc}{section}{References}
\bibliographystyle{LHCb}
\bibliography{main,LHCb-PAPER,LHCb-CONF}


\end{document}